\documentclass[11pt]{article}
\usepackage{graphicx}

\begin{document}

\def\xslash#1{{\rlap{$#1$}/}}
\def \p {\partial}
\def \dd {\psi_{u\bar dg}}
\def \ddp {\psi_{u\bar dgg}}
\def \pq {\psi_{u\bar d\bar uu}}
\def \jpsi {J/\psi}
\def \psip {\psi^\prime}
\def \to {\rightarrow}
\def\bfsig{\mbox{\boldmath$\sigma$}}
\def\bfeps{\mbox{\boldmath$\epsilon$}}
\def\DT{\mbox{\boldmath$\Delta_T $}}
\def\xit{\mbox{\boldmath$\xi_\perp $}}
\def \jpsi {J/\psi}
\def\bfej{\mbox{\boldmath$\varepsilon$}}
\def \t {\tilde}
\def\epn {\varepsilon}
\def \up {\uparrow}
\def \dn {\downarrow}
\def \da {\dagger}
\def \pn3 {\phi_{u\bar d g}}

\def \p4n {\phi_{u\bar d gg}}

\def \bx {\bar x}
\def \by {\bar y}

\begin{center}
{\Large\bf Rescattering Effect and Near Threshold Enhancement of $p\bar p$ System}
\vskip 10mm
G.Y. Chen$^1$, H.R. Dong$^2$ and J.P. Ma$^{2,3}$   \\
{\small {\it $^1$ Department of Physics, Peking University,
Beijing 100871, China }} \\
{\small {\it $^2$ Institute of Theoretical Physics, Academia Sinica,
Beijing 100080, China }} \\
{\small {\it $^3$ Theoretical Physics Center for Science Facilities, Academia Sinica,
Beijing 100049, China }} \\
\end{center}
\begin{abstract}
We study the observed enhancement of a $p\bar p$ system near the threshold
in the process $J/\psi \to \gamma p\bar p$ and $e^+ e^- \to p\bar p$. From early studies
the enhancement can be explained by final state interactions, which are in general taken into account
with some potential models. In this work
we offer a simple approach within quantum field theory to explain the observed enhancement.
We point out that among different final state interactions
the rescattering in a $N\bar N$ system though exchange of $\pi$ is the most important.
The effects of the rescattering is completely fixed by the well-known coupling $g_{\pi NN}$.
Our results show that the enhancement in $J/\psi \to \gamma p\bar p$ and $e^+ e^- \to p\bar p$
can be well described with the rescattering effects.
\end{abstract}

\par\vskip20pt
\noindent
{\bf 1. Introduction}
\par
Recently it has been reported by several experimental groups unusual enhancement of a
baryon-antibaryon system produced near the threshold. In the decay $J/\psi \to \gamma  p \bar p$
the enhancement has been observed by BES\cite{BES1} near the threshold of the $p\bar p$ system.
BES also has observed the enhancement in the decay $J/\psi \to K^- p \bar \Lambda$ near the
threshold of the $p\bar \Lambda$-system\cite{BES2}. In the decay of $B^+\to K^+ p\bar p$ and
$\bar B^0 \to D^0 p\bar p$ the enhancement has been observed by Belle near the threshold
of the $p \bar p$ system\cite{Bell}. Recently Barbar has reported the enhancement in the process
$e^+ e^- \to p\bar p, \Lambda \bar \Lambda, \Sigma^0 \bar \Sigma^0$\cite{Bar1, Bar2}, respectively.
These experimental results have stimulated many theoretical speculations\cite{Tbes1,Tbes2,Tbes3,ZC,HML,Tbell}.
It seems that the observed enhancement is a general feature for a baryon-antibayron system produced
near its threshold. In this work we focus on the enhancement of $p\bar p$ system observed at BES and BarBar.
\par
The experimental data of the observed enhancement in the decay $J/\psi\to\gamma p \bar p$ can be described
with an $S$-wave Breit-Wigner resonance function with a peak mass at $M=1859\pm 6$MeV below the threshold\cite{BES1},
and an analysis of the angular distribution of the photon suggests that the $p\bar p$ system is likely with
the total angular momentum $J=0$\cite{BES1}.
Many explanations for the observed enhancement at BES exist. A class of explanations is that
the enhancement is interpreted as the existence of a baryonium bound state\cite{Tbes1} or a glueball
below the threshold\cite{BesG}.
Another class of explanations is to take the effect of final state interactions into account. There are different
ways to take final state interactions into account. One can use a complex $S$-wave $p\bar p$ scattering length\cite{Tbes2}
or use a $K$-matrix formalism to include one pion exchange\cite{ZC}.
A more realistic way is by using models of $N\bar N$ interactions\cite{Tbes3,Tbes4}. These models are partly fixed
by a well-known $NN$ interacting potential which corresponds to the dispersive part of the $N\bar N$ interaction.
The observed enhancement in $e^+e^- \to p\bar p$ has also motivated some theoretical
studies\cite{Tbes4,Tbar1,Tbar2,Tbar3}. It is interesting to note that by taking final state interactions
into account through models of $N\bar N$ interactions, the enhancement in $J/\psi \to \gamma p\bar p$ and
$e^+e^- \to p\bar p$ can be explained\cite{Tbes4,Tbar2}. However, these models are in general complicated
and contain several or more parameters which need to be fixed.
In this work we offer a simple approach within quantum field theory to explain the observed enhancement
in the $J/\psi$ decay and $e^+e^-$ annihilation.
\par
Once a $N\bar N$ system is produced, the final state interaction can happen in several ways.
Near the threshold, the most important final state interaction is expected to be the rescattering chain process,
the rescattering can be multiple, i.e., $N\bar N \to N\bar N \cdots \to N\bar N$, where only $N\bar N$'s are
in intermediate states.
Since the momentum transfer
near the threshold is small and approaches to zero, one can expect that the most important rescattering
is through the exchange of one pion, because pion is the meson with the smallest mass. This can be seen
from the propagator of the exchanged particle. With this argument one also finds that the intermediate states
can only be  $N\bar N$ states. Otherwise, the exchanged particle is heavier than pion and its contribution
is suppressed. It is possible to have more than two particles in intermediate states, but the contributions
from these states are suppressed by phase-space factors and also by propagators of particles heavier than pions.
At the low energy, the coupling constant of $\pi NN$
is fixed as $g_{\pi NN}$. Therefore in our approach there is only one well-known parameter.
In fact such an effect
of rescattering has been considered for the $J/\psi$ decay in \cite{ZC} where the interaction
is described by a potential. Our results are different than those in \cite{ZC} and the difference
will be discussed. Within our approach
we can explain the enhancement observed in $J/\psi \to \gamma p\bar p$  and $e^+ e^- \to p\bar p$.
In our approach we will neglect the Coulomb enhancement factor, because it only has a significant effect within
a few MeV above the threshold.
\par
Our work is organized as the following: in Sect.2 we give our result for the multiple $N\bar N$ rescattering
though the $\pi NN$ interaction. The effect of the multiple rescattering can be summed in amplitudes and analytical
results for amplitudes are given. In Sec.3 we give our numerical results for the enhancement
in $J/\psi \to \gamma p\bar p$ and also for the decay of $J/\psi \to \pi^0 p\bar p$.
In Sec.4 we compare our predictions with experimental results of
$e^+e^- \to p\bar p$. Sect.5 is our summary.

\par\vskip20pt

{\noindent}
{\bf 1. Rescattering in $N\bar N$ Systems}
\par
As mentioned in the introduction we consider the rescattering of $N\bar N$ system through the interaction
between $\pi$ and $N$ at low energy. The interaction is well-known and given through the effective
Lagrangian as:
\begin{equation}
   \delta {\mathcal L} = i g_{\pi NN} \bar N \tau^i \gamma_5 N \pi^i,
   \ \ \ \ \ N = \left ( \begin{array}{cc} p \\ n \end{array} \right ),
\end{equation}
where $N$ is the field for nucleon $N$, $\pi^i(i=1,2,3)$ is the pion field and $\tau^{i}(i=1,2,3)$
is the Pauli matrix acting in the isospin space. For a simple representation of our approach
and results we first study the rescattering process of single channel, i.e.,
$p\bar p\to p\bar p \cdots \to p\bar p$.
Then we generalize the results of the single channel to those of a $N\bar N$ system.
\par
We consider a $p\bar p$ system produced through a vertex $\Gamma$, then the rescattering
of the system can happen through the exchange of $\pi^0$, as drawing in Fig.1, where the black
circle represents the vertex $\Gamma$. At first look it may be meaningless to sperate the production
amplitude into a vertex part and a rescattering part, because the combination of the two parts
is exactly the vertex for the production. However the separation is still meaningful if one takes
the vertex in Fig.1 as that obtained from a analytical continuation of the vertex in the space-like region
into the time-like region above the threshold, where only the dispersive part is taken near the threshold.
In that sense the absorptive part of the production amplitude comes only from the rescattering. We will call
the vertex as "bar" vertex.

\par
\begin{figure}[hbt]
\begin{center}
\includegraphics[width=10cm]{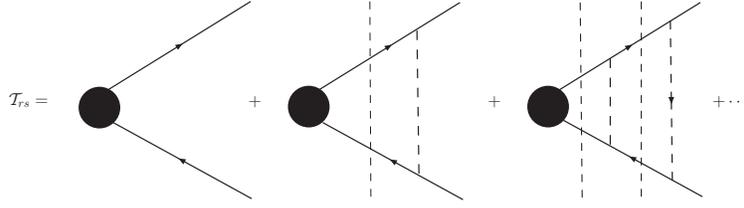}
\end{center}
\caption{The Feynman diagrams for the rescattering between a proton
and an antiproton through the exchange of $\pi^0$. The dash lines cutting
diagrams are cuts.} \label{Feynman-dg1}
\end{figure}
\par
The production amplitude in our approach is then the sum
of the amplitudes in which exchanges of arbitrary number of $\pi^0$ happen.
We denote the amplitude the rescattering though  $n$-$\pi^0$ exchange as ${\mathcal T}_n$. The amplitude
without the exchange is denoted as ${\mathcal T}_0$, which is given by
\begin{equation}
   {\mathcal T}_0 = \bar u(p_1,s_1) \Gamma^{(p)} (p_1,p_2) v(p_2,s_2).
\end{equation}
In the above $\Gamma$ is the "bar" vertex. The proton carries the momentum $p_1$ and the spin $s_1$
and the antiproton carries the momentum $p_2$ and the spin $s_2$.
The amplitude with the rescattering through one $\pi^0$ can be obtained with Cutkosky rules as:
\begin{eqnarray}
{\mathcal T}_1 &=& 2 i g_{\pi NN}^2 \pi^2 \int \frac{ d^4 k_1}{(2\pi)^4} \delta (k_1^2 -m_p^2) \delta (k_2^2 -m_p^2)
                     \cdot \frac{1}{ (k_1 -p_1)^2 -m_\pi^2 }
\nonumber\\
    && \cdot \sum_{s,s'} \bar u(p_1, s_1) \gamma_5 u(k_1,s) \bar u (k_1,s) \Gamma^{(p)} (k_1,k_2) v(k_2,s')
         \bar v(k_2,s') \gamma_5 v(p_2,s_2),
\nonumber\\
     k_2 &=& p_1+p_2 -k_1.
\end{eqnarray}
Similarly, one can write down the amplitude ${\mathcal T}_n$. The amplitude with all rescattering effects
is the sum:
\begin{equation}
  {\mathcal T}_{rs} = \sum_{n=0}^{\infty} {\mathcal T}_n.
\end{equation}
\par
We will work with the rest frame of the $p\bar p$ system.
In the system the momenta are:
\begin{equation}
  p_1^\mu =(E, {\bf p}), \ \ \ \ \  p_2^\mu = (E, -{\bf p}).
\end{equation}
We assume that the dependence of the production rate on the invariant mass $q^2$ with $q^2=(p_1+p_2)^2$
near the threshold is dominantly determined by the final state interaction or rescattering. With
the assumption one can expand the vertex and the products of spinors in the small ${\bf p}$:
\begin{eqnarray}
\Gamma^{(p)} (p_1,p_2) & = & \Gamma^{(p)} (E,{\bf p}) =  \Gamma^{(p)} (m_p, 0) \left \{ 1 + {\mathcal O}(\beta )\right \},
\nonumber\\
  \bar u (p_1,s_1) \gamma_5 u(k_1,s) &=& \xi^\dagger (s_1) \bfsig \cdot ({\bf k_1} -{\bf p_1}) \xi(s)
    \left \{ 1 + {\mathcal O}(\beta )\right \},
\nonumber\\
 \bar v(k_2,s) \gamma_5 v(p_2,s_2) &=& \eta^\dagger (s)  \bfsig \cdot ({\bf k_2} -{\bf  p_2}) \eta (s_2)
   \left \{ 1 + {\mathcal O}(\beta )\right \},
\nonumber\\
   \beta &=& \sqrt { 1 -\frac{4m_p^2}{q^2}}=\frac{2}{\sqrt{q^2}}\vert {\bf p}\vert , \ \ \ \  q^2 =(p_1+p_2)^2,
\end{eqnarray}
where $\xi$ is the two-component spinor for the proton and $\eta$ is that for the anti-proton.
The above expansion implies that we take the nonrelativistic limit $\beta \to 0$.
In the limit we can classify the state of system with the sum of spins $s=s_1+s_2$ by using:
\begin{eqnarray}
   \xi (s_1) \eta^\dagger (s_2) \langle s_1,s_2 \vert s=0,s_z=0 \rangle &=& \frac{1}{\sqrt{2}} I,
\nonumber\\
   \xi (s_1) \eta^\dagger (s_2) \langle s_1,s_2 \vert s=1,s_z \rangle &=& \frac{1}{\sqrt{2}}
    \bfsig \cdot \bfeps (s_z),
\end{eqnarray}
where $I$ is a $2\times 2$ unit matrix, $\langle s_1,s_2 \vert S,S_z\rangle$ is the Clebsch-Gordan
coefficient and $\bfeps (s_z)$ is the polarization vector for the case $s=1$. We can decompose
the amplitude into a $s=0$-part and a $s=1$ part:
\begin{eqnarray}
  {\mathcal T}_1 &=& {\mathcal T}_1^{s=0} + {\mathcal T_1}^{s=1},
\nonumber\\
  {\mathcal T}_1^{s=0} &=& i g_{\pi NN}^2 \pi^2 \int \frac{ d^4 k_1}{(2\pi)^4} \delta (k_1^2 -m_p^2) \delta (k_2^2 -m_p^2)
                     \cdot \frac{1}{ (k_1 -p_1)^2 -m_\pi^2 }
\nonumber\\
       && \cdot  {\rm Tr} \left [ \bfsig \cdot ({\bf k_1} -{\bf p_1}) \bfsig \cdot({\bf p_1} -{\bf k_1})
       \right ] {\mathcal A}_{p\bar p} ,
\nonumber\\
     {\mathcal T}_1^{s=1} &=& i g_{\pi NN}^2 \pi^2 \int \frac{ d^4 k_1}{(2\pi)^4} \delta (k_1^2 -m_p^2) \delta (k_2^2 -m_p^2)
                     \cdot \frac{1}{ (k_1 -p_1)^2 -m_\pi^2 }
\nonumber\\
       && \cdot {\rm Tr} \left [  \bfsig \cdot \bfeps^* (s_z)
       \bfsig \cdot({\bf k_1} -{\bf p_1})\bfsig \cdot \bfeps (s_z') \bfsig \cdot({\bf p_1} -{\bf k_1})
       \right ] {\mathcal B}_{p\bar p}(s_z'),
\nonumber\\
     {\mathcal A}_{p\bar p} &=&  \langle 0,0 \vert s_1,s_2 \rangle \bar u(p_0,s_1)\Gamma^{(p)}(m_p,0) v(p_0, s_2),
\nonumber\\
      {\mathcal B}_{p\bar p}(s_z) &=& \langle 1,s_z \vert s_1,s_2 \rangle \bar u(p_0,s_1)\Gamma^{(p)} (m_p,0) v(p_0, s_2).
\end{eqnarray}
The summation over repeated spin indices is implied. The amplitude ${\mathcal A}_{p\bar p}$ is the amplitude
for the $p\bar p$ in the state $^1S_0$ before the rescattering, the amplitude ${\mathcal B}_{p\bar p}(s_z)$
is that for the $p\bar p$ in the state $^3S_1$ before the rescattering.
\par
It is straightforward to obtain the result of ${\mathcal T}_1$ by performing
the phase-space integral.
We have for the $s=0$ part:
\begin{equation}
{\mathcal T}_1^{s=0} =  i d(q^2) {\mathcal A}_{p\bar p},\ \ \ \
  d (q^2) = \frac{g_{\pi NN}^2 \beta}{32 \pi} \left ( 2
   - y \ln \left (1 + \frac{2}{y} \right ) \right ), \ \ \ \ \ y =\frac{2 m^2_\pi }{q^2 \beta^2}.
\end{equation}
With arguments of symmetry one can show that the total spin $s$ will not be changed after multiple scattering.
Hence we have:
\begin{equation}
 {\mathcal T}_{rs}={\mathcal T}_{rs}^{s=0} +{\mathcal T}_{rs}^{s=1}, \ \ \
 {\mathcal T}_{rs}^{s=0}=\sum_n {\mathcal T}_n^{s=0}, \ \ \ \ {\mathcal T}_{rs}^{s=1}=\sum_n {\mathcal T}_n^{s=1}.
\end{equation}
Inspecting the structure of the amplitude ${\mathcal T}^{s=0}$ one easily finds the result and the sum:
\begin{equation}
 {\mathcal T}_n^{s=0} = \left [  i d(q^2) \right ]^n
   {\mathcal A}_{p\bar p}, \ \ \ \ \
{\mathcal T}_{rs}^{s=0} = \sum_{n=0}^{\infty} {\mathcal T}_n^{s=0}=
\frac{{\mathcal A}_{p\bar p}}
{ 1 - i d(q^2) }.
\end{equation}
From the above expressions one can see that the $q^2$-dependence in the amplitude
appears through the variable $y$ combined with $m_\pi$ in Eq.(9). Hence the energy scale $m_\pi$ characterizes
the $q^2$-dependence. In the limit $\beta \to 0$ the amplitude is proportional to
$m^{-2}_\pi$. If other particles are exchanged, then instead of $m_\pi$ their masses
appear in $y$ and characterize the $q^2$-dependence. Since these particles must have masses $M$ larger than
$m_\pi$, their
contributions to the amplitude will be suppressed by $m^{2}_\pi/M^2$ in comparison with that of $\pi$-exchange
and lead to a small correction to the $q^2$-dependence in the above.
This is the argument given in the introduction to support our approach.
\par
Now we turn to the $s=1$ part. In our approach, the $p\bar p$ system produced from the vertex without scattering
has the orbital angular momentum $\ell =0$. After the rescattering with the interaction in Eq.(1), the orbital angular
momentum can be $\ell =0$ and $\ell=2$. The total angular momentum $J$ remains the same as $J=1$.
With the rescattering the amplitude will have two components,
one is with $\ell=0$, another is with $\ell=2$. The two components are at the same order of $\beta$.
We introduce the notation for the two components:
\begin{equation}
  (p\bar p)_{0}(s_z,m_z)=\bfeps^* (s_z) \cdot \bfeps (m_z), \ \ \ \
  (p\bar p)_{2}(s_z,m_z)={\bf\hat p} \cdot \bfeps^*(s_z) {\bf\hat p} \cdot \bfeps(m_z)
       -\frac{1}{3}\bfeps^* (s_z) \cdot \bfeps (m_z).
\end{equation}
With the notation we can present the result for ${\mathcal T}_1$ as:
\begin{equation}
{\mathcal T}_1^{s=1} =  {\mathcal B}_{p\bar p}(m_z)  \left ( 1, 0 \right ) \left [ i\frac {\beta g_{\pi NN}^2}{64\pi } {\mathcal M } \right ]
      \left(\begin{array}{cc} (p\bar p)_{0}(s_z,m_z)  \\
       (p\bar p)_{2}(s_z,m_z)
       \end{array}\right),
\end{equation}
where ${\mathcal M}$ is a $2\times 2$ matrix given by:
\begin{eqnarray}
{\mathcal M} &=& \left ( \begin{array}{cc} {\mathcal M}_{00}, &  {\mathcal M}_{02} \\
                          {\mathcal M}_{20}, & {\mathcal M}_{22} \end{array} \right )
 = \left(\begin{array}{cc} -\frac{2}{3}  ( 2 -y L_y ), &  ( 2 -6y +(3y^2+2y) L_y)  \\
       \frac{2}{3} \left ( \frac{2}{3} -2y + \left ( y^2 +\frac{2y}{3}\right ) L_y \right ),&
       \left ( -\frac{2}{3} -2y +\left (y^2 +\frac{4y}{3} \right)  L_y \right )
       \end{array}\right),
\nonumber\\
  L_y &=& \ln\left (1 +\frac{2}{y} \right ).
\end{eqnarray}
In the above the row vector $(1,0)$ indicates that the $p\bar p$ pair produced by the vertex
is with $\ell =0$. The matrix elements ${\mathcal M}_{20}$ and ${\mathcal M}_{22}$ are actually
identified through the calculation of ${\mathcal T}_2$. They will not contribute to ${\mathcal T}_1$
here. The above result also indicates that the rescattering of the $p\bar p$ pair with $s=1$ is already a
coupled channel problem.
Again, the structure of ${\mathcal T}_n^{s=1}$ can be easily found and the sum can be obtained:
\begin{eqnarray}
{\mathcal T}_n^{s=1} &=& {\mathcal B}_{p\bar p}(m_z)
   \left ( 1, 0 \right ) \left [ i\frac {\beta g_{\pi NN} ^2}{64\pi} {\mathcal M } \right ]^n
      \left(\begin{array}{cc}  (p\bar p)_{0}(s_z,m_z) \\
       (p\bar p)_{2}(s_z,m_z)
       \end{array}\right),
\nonumber\\
{\mathcal T}_{rs}^{s=1} &=& \sum_{n=0}{\mathcal T}_n^{s=1}
         = {\mathcal B}_{p\bar p}(m_z) (1,0) \left [ I - i\frac {\beta g_{\pi NN}^2}{64\pi } {\mathcal M } \right ]^{-1}
     \left(\begin{array}{cc}  (p\bar p)_{0}(s_z,m_z) \\
       (p\bar p)_{2}(s_z,m_z)
       \end{array}\right).
\end{eqnarray}
\par
The above results can be generalized to the rescattering of a $N\bar N$ system with the given interaction
by taking the isospin factor into account. We will neglect the mass difference between protons
and neutrons and that between different $\pi$'s. These differences are small and only give
small corrections to our results.
For ${\mathcal T}_1$ the isospin factor can be determined by considering the scattering
$N_i \bar N_k \to N_j \bar N_l$ through one-$\pi$ exchange.
The isospin factor in the amplitude is:
\begin{equation}
 (\tau^a)_{ji} (\tau^a)_{kl} = 2 \delta_{jl}\delta_{ik} -\delta_{ji} \delta_{kl}.
\end{equation}
We denote the "bar" vertex for $p\bar p$ and $n\bar n$ as $\Gamma^{(p)}$ and $\Gamma^{(n)}$, respectively.
We have for the $s=0$ part:
\begin{eqnarray}
 {\mathcal T}_n^{s=0} &=&  \left ({\mathcal A}_{p\bar p}, {\mathcal A}_{n\bar n} \right )
   \left [  i d(q^2) {\mathcal C} \right ]^n
 \left(\begin{array}{cc}  (p\bar p) \\
       (n\bar n)
       \end{array}\right),
\nonumber\\
{\mathcal T}_{rs}^{s=0} &=&  \sum_{n=0}^{\infty} {\mathcal T}_n^{s=0}=
 \left ({\mathcal A}_{p\bar p}, {\mathcal A}_{n\bar n} \right ) \left [ I  -  i d(q^2) {\mathcal C} \right ]^{-1}
 \left(\begin{array}{cc}  (p\bar p) \\
       (n\bar n)
       \end{array}\right),
\end{eqnarray}
with
\begin{equation}
{\mathcal C} = \left (\begin{array}{cc} 1, &2\\ 2,& 1 \end{array} \right ).
\end{equation}
${\mathcal C}$ is a $2\times 2$ matrix acting in the isospin space. $(p\bar p)$ and $(n\bar n)$ denotes
the $^1S_0$ states of $p\bar p$ and $n\bar n$, respectively. ${\mathcal A}_{n\bar n}$ is obtained
by replacing $\Gamma^{(p)}$ in ${\mathcal A}_{p\bar p}$ with $\Gamma^{(n)}$. We note for the above result
that the matrix ${\mathcal C}$ is diagonal if we use the basis of isospin.
\par
For the $s=1$ part the results are:
\begin{eqnarray}
{\mathcal T}_n^{s=1} &=& \left ( {\mathcal B}_{p\bar p} (m_z), 0,  {\mathcal B}_{n\bar n} (m_z), 0 \right )
     \left [ + i\frac {g^2_{\pi NN} \beta }{ 64\pi } {\mathcal C}\otimes {\mathcal M }\right ]^n
      \left(\begin{array}{cc}  (p\bar p)_{0}(s_z,m_z) \\
       (p\bar p)_{2}(s_z,m_z)\\ (n\bar n)_{0}(s_z,m_z) \\
       (n\bar n)_{2}(s_z,m_z)
       \end{array}\right),
\nonumber\\
{\mathcal T}_{rs}^{s=1} &=& \sum_{n=0}{\mathcal T}_n^{s=1}
         = \left ( {\mathcal B}_{p\bar p} (m_z), 0,  {\mathcal B}_{n\bar n} (m_z), 0 \right )
          \left [ E - i\frac {\beta g_{\pi NN} ^2}{64\pi}  {\mathcal C}\otimes {\mathcal M }\right ]^{-1}
     \left(\begin{array}{cc}  (p\bar p)_{0}(s_z,m_z) \\
       (p\bar p)_{2}(s_z,m_z)\\ (n\bar n)_{0}(s_z,m_z) \\
       (n\bar n)_{2}(s_z,m_z)
       \end{array}\right),
\end{eqnarray}
where $E$ is a $4\times 4$ unit matrix. Now we can use the above results to make predictions
for relevant experiments.
\par
Before turning to our numerical predictions, it is important to know the $q$-region where our approach
is applicable.
In our approach we have considered the rescattering effect among a $N\bar N$ system which contains the two most lightest
baryons. The next lightest baryon is $\Lambda$. In the range of $q^2 > 4 m^2_\Lambda$ $\Lambda$'s will enter into
the rescattering as intermediate states which is not included in our results.
Therefore our result may not be useful for $q^2>4 m^2_\Lambda =
(2\cdot 1115 {\rm MeV})^2$.
We have also taken the nonrelativistic limit $\beta \to 0$ in our approach.
The effect of the next-to-leading order in $\beta$ can be important if $\beta$ is large.
For ${\sqrt{q^2}-2m_p}\sim 200{\rm MeV}$
$\beta^2$ takes the value $\sim 0.2$ which is not large. Therefore we expect that our results should be applicable
in the range ${\sqrt{q^2}-2m_p} <100\sim  200{\rm MeV}$. It should be noted that
it is possible to include higher-order effects in $\beta$ and other neglected effects discussed in the above.
\par

\par\vskip20pt
\noindent
{\bf 3. The Enhancement in the Decay $J/\psi \to \gamma p\bar p$}
\par
There is an enhancement of the $p\bar p$ system near threshold in the decay $J/\psi \to \gamma p \bar p$
observed by BES and the experimental results favor that the $p\bar p$ pair near the threshold
is in the state $^1S_0$. This corresponds to our result with $s=0$.
In the experimental result of BES\cite{BES1}, most data points lie in the region with $q^2 > 4m_n^2$.
Therefore $n\bar n$ pairs can appear
in the intermediate states.
It is expected that not only a $p\bar p$ but also a $n\bar n$ can be produced
through the "bar" vertex
in the radiative decay of $J/\psi$.
From Eq. (17) we obtain the decay amplitude with the rescattering effect for  $q^2 >4 m_n^2$:
\begin{equation}
{\mathcal T}_{rs}^{s=0} = \frac { {\mathcal A}_{p\bar p} + {\mathcal A}_{n\bar n} }{2}
\cdot \frac{1 + id(q^2)}{(1-id (q^2))^2 +4 d^2 (q^2) } +
\frac{{\mathcal A}_{p\bar p} -{\mathcal A}_{n\bar n} }{2}
\cdot \frac{1 -3id(q^2)}{(1-id (q^2))^2 +4 d^2 (q^2) },
\end{equation}
where the first term is with the isospin $I=0$ and the second is with $I=1$.
The constants ${\mathcal A}_{p\bar p,n\bar n}$ are unknown because the isospin in the decay
is violated in general because of the electromagnetic interaction.
In our numerical predictions we will take the constants ${\mathcal A}_{p\bar p, n\bar n}$
as free parameters to fit the BES results.
\par
\par
\begin{figure}[hbt]
\begin{center}
\includegraphics[width=11cm]{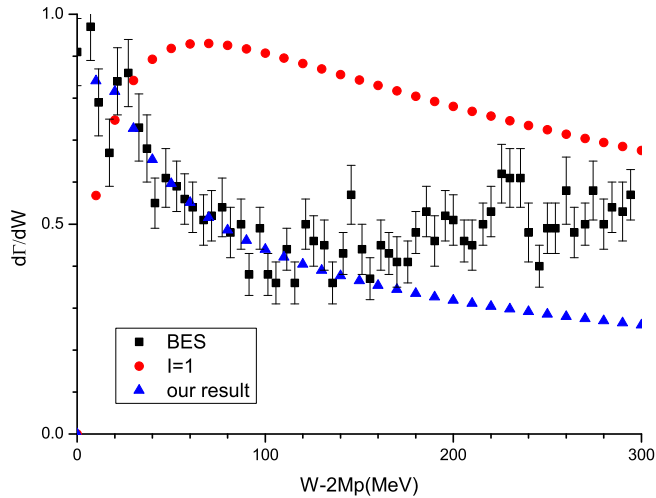}
\end{center}
\caption{Comparison of BES results $J/\psi \to \gamma p\bar p$ with our predictions. }
\label{Feynman-dg1}
\end{figure}
\par
In Fig.2 we draw the BES date from \cite{BES1} and our fitting results.
We have fitted the experimental date in the region with $\sqrt{q^2}-2m_p<150$MeV. The fitted parameters
are ${\mathcal A}_{p\bar p}\approx 0.895$ and ${\mathcal A}_{n\bar n}\approx 1.027$ with
$\chi^2/{\rm d.o.f.} \approx 1.14$.
From Fig.2 we can see that our predicted $q^2$-dependence in Eq.(20) matches fairly well
with experimental data  in the region with 150MeV$>\sqrt{q^2}-2m_p$.
Our prediction fails to describe the experimental data in the region with $\sqrt{q^2}-2m_p>150$MeV.
The reason can be that the higher-order effects in $\beta$ or other neglected effects can become important
in this case. However, from Fig.2 we see that the observed enhancement happens in the region
with $\sqrt{q^2}-2m_p <150$MeV, we can conclude that the enhancement is fairly well in agreement
with our result. Also, our results indicate that the $p\bar p$ near the threshold
is almost in the state with $I=0$ because
${\mathcal A}_{n\bar n}+{\mathcal A}_{p\bar p} >> {\mathcal A}_{n\bar n}-{\mathcal A}_{p\bar p}$
from our fitting results.
\par
\begin{figure}[hbt]
\begin{center}
\includegraphics[width=11cm]{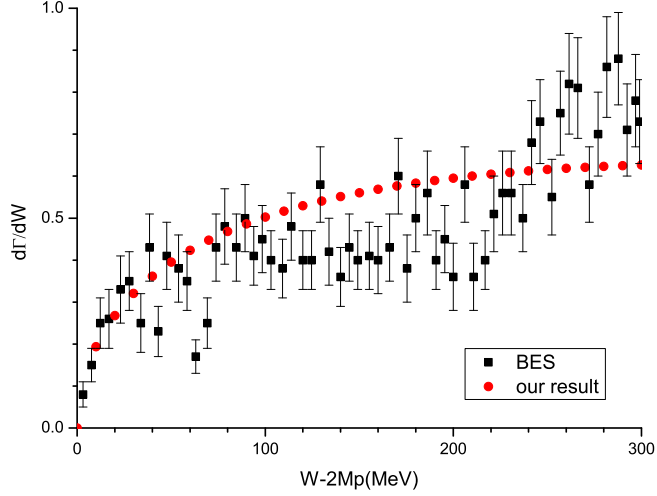}
\end{center}
\caption{Comparison of BES results of $J/\psi \to \pi^0 p\bar p$ with our predictions. }
\label{Feynman-dg1}
\end{figure}
\par
In \cite{ZC} the rescattering effect in the $I=0$ channel through $\pi$-exchange
are studied where the interaction is given by a potential. We note that the potential used
in \cite{ZC} is different than the potential derived from the interaction Lagrangian $\delta {\mathcal L}$
in Eq.(1). This leads to the difference between our predictions and those in \cite{ZC}.
Numerically, the results for the spectrum obtained in \cite{ZC} look very similar to ours in the
$I=1$ channel, which we also give in Fig.2.
From Fig.2. it is clearly that the enhancement can not be explained only with the
rescattering effect in the $I=1$ channel. Our result is also different than
that in \cite{Tbes3}, in which the enhancement is explained with the $p\bar p$ system
in the $I=1$ state.
\par
In experiment the decay $J/\psi \to \pi^0 p\bar p$ has been studied, but no enhancement near the threshold
of the $p\bar p$ system has been observed\cite{BES1}. It has been a puzzle.
In general the final state interaction or rescattering
is more complicated than the radiative decay, because one
has three hadrons in the final state and the final state interaction exists
not only in the $p\bar p$ system but also between the system and $\pi^0$.
If we assume that the rescattering effect of the $N\bar N$ system
is dominant, then the observation can also be explained in our approach.
In the decay the $p\bar p$ system is in the $I=1$ state. Near the threshold
the $p\bar p$ system can only be in a $^3S_1$ state
or a $^1 P_1$ state. For the $P$-wave state the decay amplitude
is suppressed by an extra $\beta$ in comparison with that of a $S$-wave state.
If we take the $p\bar p$ system produced through the "bar" vertex as a $^3S_1$ state,
we can give predictions in our approach. In Fig.3 we give our numerical
results from Eq.(19) in comparison with BES data. It can be seen that our approach
can explain the $q^2$-dependence fairly well near the threshold.
\par
Recently, BES has reported that no enhancement in the decay $J/\psi \to \omega p\bar p$
has been found near the threshold\cite{BesOm}. In this decay the $p\bar p$ system
is in the state with $I=0$. If one assumes that the rescattering effect of the $N\bar N$ system
is dominant as in the above, one expects from our approach that there will be
an enhancement. The existence of the enhancement is also
predicted in \cite{HMS}. However, the assumption may not be correct. In general, for a complete
prediction in this case one needs
to work out rescattering effects among three hadrons. This deserves a further study.
\par\vskip20pt
\noindent
{\bf 4. The Enhancement in $e^+ e^- \to p\bar p$}
\par
The total cross section of $e^+ e^- \to \gamma^* \to p \bar p$ can be expressed in terms of form factors of protons:
\begin{equation}
\sigma (e^+ e^- \to p\bar p ) = \frac {4 \pi \alpha^2 \beta  } {3 q^2} \left \{
    \vert G_M^{(p)} \vert ^2 + \frac{ 2  m_p^2}{q^2}    \vert G_E^{(p)} \vert ^2 \right \},
\end{equation}
where $G_M$ and $G_E$ are the Sachs form factors of protons. They are related with the Dirac- and Pauli
form factor. The definitions of these form factors are:
\begin{eqnarray}
 \langle p(p_1) \bar p (p_2) \vert J^\mu \vert 0\rangle & = &  \bar u(p_1) \left [ \gamma^\mu F_1 (q^2)
      + i\frac{\sigma^{\mu\alpha}}{2 M} q_\alpha F_2(q^2) \right  ] v(p_2),
\nonumber\\
   G_M &=&  F_1 + F_2, \ \ \ \  G_E = F_1 + \frac{q^2}{4 M^2} F_2, \ \ \ \  q=p_1+p_2.
\end{eqnarray}
In our approach the measured form factors are related to the "bar" form factors
through rescattering.
Since we are interested in the behavior near the threshold,
we expand the above spinors in $\beta$.
It should be noted that with our formula given in Eq. (19) it is possible to keep the
relativistic effect from the spinors. Including this effect we have for the matrix element:
\begin{eqnarray}
\langle p(p_1) \bar p (p_2) \vert {\bf J } \vert 0\rangle & = & -2M \left ( F_1(q^2)+ F_2(q^2) \right ) \left [
 \left (1+ \frac{\vert {\bf p}\vert^2}{3M^2} \right ) \xi^\dagger \bfsig \eta
  -\frac{\vert {\bf p}\vert^2}{2M^2} \xi^\dagger \left [ \bfsig\cdot {\bf\hat p} {\bf\hat p} - \frac{1}{3} \bfsig \right ]
  \eta \right ]
\nonumber\\
  && - 2 F_2(q^2)  \frac{\vert {\bf p}\vert^2}{M}\xi^\dagger
    \left [ \bfsig\cdot {\bf\hat p} {\bf\hat p} - \frac{1}{3} \bfsig \right ] \eta
      -2 F_2 (q^2) \frac{\vert {\bf p}\vert^2}{3 M} \xi^\dagger \bfsig \eta +\cdots,
\end{eqnarray}
where $\cdots$ stand for contributions at $\beta^4$ and are neglected in our numerical predictions.
We denote the form factors in the "bar" vertex, i.e., the "bar" form factors, as $\tilde F_{1,2}$
and the inverse of the $4\times 4$ matrix in Eq.(19) as:
\begin{equation}
     {\mathcal A} =   \left [ I - i\frac {\beta g_{\pi NN} ^2}{32(2\pi)^2 } {\mathcal C}\otimes {\mathcal M }\right ]^{-1},
\end{equation}
then we have the form factors of protons with the rescattering effect
near the threshold as:
\begin{eqnarray}
\left [( F_1(q^2) +F_2(q^2) ) \left ( 1 + \frac{\vert {\bf p}\vert^2}{3M^2} \right )
                  + F_2(q^2) \frac{ \vert {\bf p}\vert^2}{3 M^2}  \right ]^{(p)}
& = & {\mathcal A}_{11}  \left (\tilde F_1 + \tilde F_2 \right  )^{(p)}
   +{\mathcal A}_{31}  \left (\tilde F_1 + \tilde F_2 \right  )^{(n)},
\nonumber\\
  -\frac{  \vert {\bf p}\vert^2}{2 M^2} \left ( F_1(q^2) -F_2(q^2) \right )^{(p)}
     &=& {\mathcal A}_{12}  \left (\tilde F_1 + \tilde F_2 \right  )^{(p)}
   +{\mathcal A}_{32}  \left (\tilde F_1 + \tilde F_2 \right  )^{(n)},
\end{eqnarray}
where the last equation comes from the $\ell =2$ part. The indices $(p)$ and $(n)$ denote
quantities for proton and neutron, respectively. In the above the "bar" form factors
should be taken with $q^2=4m_p^2$ in our approach, see Eq.(6). Therefore,  the $q^2$-dependence
of the form factors and of the cross-section comes from the rescattering effect and
the relativistic correction in the expansion of spinors.
It should be noted that there will be only one form factor if the relativistic effect is neglected, i.e.,
$G_E= G_M$.
\par
\begin{figure}[hbt]
\begin{center}
\includegraphics[width=11cm]{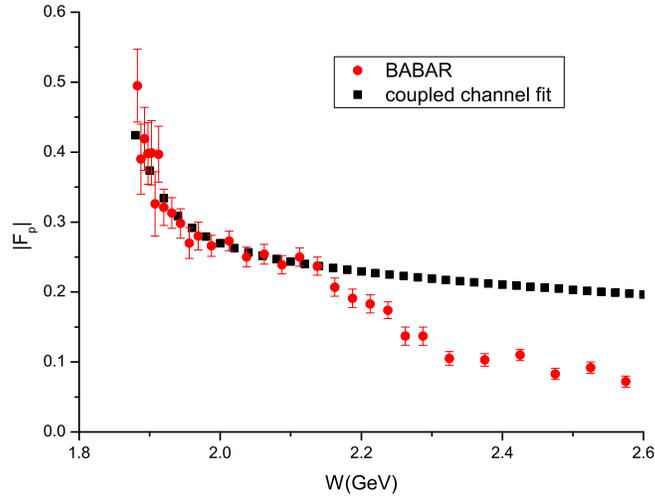}
\end{center}
\caption{Comparison of Barbar results with our predictions. }
\label{Feynman-dg3}
\end{figure}
\par
\par
With the above results one can obtain predictions for form factors and the cross-section
near the threshold. If one knows the "bar" form factors, the form factors are completely determined.
As discussed in Sect.2., these "bar" form factors should be obtained as an analytical continuation
of form factors
from the space-like region with $q^2<0$ to the time-like region with $q^2=4m_p^2$. There are intensive
studies of experiments and in theory for the form factors in space-like regions. However, the continuation
with existing results may be problematic, since in the continuation $q^2$ will cross the unphysical region
$0< q^2 < 4m_p^2$. We will take the combination $\tilde G_M^{(n)}=\left (\tilde F_1 + \tilde F_2 \right  )^{(n)}$ and
$ \tilde G_M^{(p)} = \left (\tilde F_1 + \tilde F_2 \right  )^{(p)}$ as two free parameters to fit the Barbar results.
To see the enhancement clearly, one can use the effective form factor which
is measured in \cite{Bar1}:
\begin{equation}
\vert F_p \vert^2 =  \frac{\sigma( e^+ e^- \to p\bar p)}{ \sigma_0},  \ \ \ \
   \sigma_0 = \frac {4 \pi \alpha^2 \beta  } {3 q^2} \left (
     1 + \frac{ 2  m_p^2}{q^2}  \right ) .
\end{equation}
In the Fig3.  we give our numerical results where the Barbar results in \cite{Bar1} are also plotted.
Our results are obtained by fitting the BarBar data in the region with $q^2<(2.230{\rm GeV})^2$ with
the fitting results:
\begin{equation}
 \tilde G_M^{(n)} \approx 0.213, \ \ \ \  \tilde G_M^{(p)} \approx 0.427, \ \ \ \  \chi^2/{\rm d.o.f.} \approx 1.02.
\end{equation}
From Fig.4 we can see that the behavior of the enhancement is well described with our results.
From our result given in Fig.3 for $J/\psi \to \pi^0 p\bar p$, where the $p\bar p$ is with $I=1$, one
may expect that the enhancement in $e^+ e^- \to p\bar p$ is mainly due the enhancement
in the channel with $I=0$.
But from our results we can see that a substantial $I=1$ component is needed to match
the experimental results.

\par\vskip20pt
\noindent
{\bf 5. Summary}
\par
In this work we have studied the enhancement of a $p\bar p$ system near the threshold
in the process $J/\psi \to \gamma p\bar p$ and $e^+ e^- \to p\bar p$. From early studies
the enhancement can be explained by final state interactions, which are in general taken into account
with some potential models.
We have given arguments to support that among these interactions
the rescattering effect of $N\bar N$ system though exchange of $\pi$ is the most important.
Hence we have proposed a simple approach in the framework of quantum field theory with the well-known
$\pi NN$ interaction to explain the enhancement, where the rescattering effect
is completely fixed by the well-known coupling $g_{\pi NN}$. It turns out
that the enhancement in $J/\psi \to \gamma p\bar p$ and $e^+ e^- \to p\bar p$
can be well described within our approach.
In these two cases the final state strong interaction can only happen between
the $N\bar N$ system.
With the assumption that only the rescattering effect in the $N\bar N$ system
is dominant in $J/\psi\to \pi^0 p\bar p$, we can also explain why there is no enhancement
in the decay.
\par
We have only studied in detail the enhancement with the final state which contains
a $p\bar p$ pair as hadrons. For other cases like the enhancement
in $B$-decay and the annihilation of $e^+e^-$ into baryon pairs other than
a $p\bar p$ pair, extensions of our approach are needed. This will be studied
in a future work.

\vskip 5mm
\par\noindent
{\bf\large Acknowledgments}
\par
The authors would like to thank Prof. Chao-Hsi Chang., S. Jin, B.S. Zou and H.Q. Zheng for helpful discussions. This
work is supported by National Nature Science Foundation of P.R. China((No. 10721063).
\par\vskip20pt

\par\vfil\eject

\end{document}